\documentclass[twocolumn,showpacs,preprintnumbers,amsmath,amssymb]{revtex4}

\usepackage{graphicx}
\usepackage{dcolumn}
\usepackage{bm}
\usepackage{graphics}
\usepackage{amsmath}
\usepackage{amssymb}
\usepackage{amscd}
\usepackage{afterpage}
\usepackage{float,times}
\usepackage{subfigure}
\usepackage{rotating}
\usepackage{multirow}
\usepackage{fancyheadings}
\usepackage{epsfig}
\usepackage{theorem}
\usepackage{moreverb}
\usepackage{euscript}
\usepackage{psfrag}

\begin{document}

\title{Orbifold Reduction Of The Quark-Lepton Symmetric Model}

\author{K. L. McDonald}\email{k.mcdonald@physics.unimelb.edu.au}

\author{B. H. J. McKellar}%
 \email{b.mckellar@physics.unimelb.edu.au}
\affiliation{%
School of Physics, Research Centre for High Energy Physics, The
University of Melbourne, Victoria, 3010, Australia\\
}%

\date{\today}

\begin{abstract}
We investigate the quark-lepton symmetric gauge group in five dimensions,
with the gauge symmetry broken by a combination of orbifold
compactification of the extra dimension and the Higgs mechanism. The
gauge sector of the model is investigated and contrasted with the four
dimensional case. We obtain lower bounds on the mass of the exotic
gauge bosons, the inverse compactification scale and
the exotic leptons. Light neutrinos are
obtained without requiring any scale larger than a TeV. However an ultra-violet
cut-off of order $10^{11}$~GeV is required to
suppress proton
decay inducing non-renormalizable operators. 
\end{abstract}

\maketitle
\section{Introduction\label{sec:orb_ql_intro_intro}}
Quark-lepton (QL) symmetric models, by definition, possess a discrete
symmetry whereby one interchanges quarks and leptons in the
Lagrangian. Observed low energy phenomena indicate three important
differences between quarks and leptons which must be overcome if one
is to define such a symmetry. Namely:\\
$i)$ Quarks display a threefold degeneracy corresponding to the colour degree of freedom ----- leptons possess no such degeneracy.\\
$ii)$ The masses of quarks and leptons are distinct and in particular no relationships of the type $m_e=m_u$ or $m_e=m_d$ are observed.\\
$iii)$ Quarks possess fractional electric charges, whilst leptons are integrally charged.

The standard model (SM) of particle physics has demonstrated that the
absence of obvious symmetry in low energy phenomena need not imply a
corresponding absence in the underlying theory. Spontaneous symmetry
breaking has provided a means by which one may construct theories
possessing symmetries which are not manifest in the low energy
regime.

The construction of a QL symmetric model requires one to
attribute the low energy differences between quarks and leptons to the
symmetry breaking mechanisms employed in the more symmetric underlying
theory. Thus in a QL symmetric theory leptons are expected
to come in three leptonic colours, analogous to the three colours of
quarks. The symmetry breaking of leptonic colour must then ensure that
only one of the three lepton colours remains massless until
electroweak symmetry breaking occurs. The different masses and
electric charges of quarks and leptons should also arise through the
symmetry breaking mechanism.

Previous
works~\cite{Foot:1990dw,Foot:1990um,Foot:1991fk,Levin:1993sq,Shaw:1994zs,Foot:1995xx}
have shown that it is possible to construct models which overcome the
differences between quarks and leptons listed above, thereby allowing
nature to realize a discrete QL symmetry at high
energies. Subsequent symmetry breaking as the universe cools results
in a low energy theory indistinguishable from the SM. For other recent
works containing leptonic colour
see~\cite{Babu:2003nw,Chen:2004jz,Demaria:2005gk,Demaria:2006uu,Foot:2006ie}.

Studies of QL symmetric models performed to date have
employed the Higgs mechanism to break the high energy theory down to
the SM. The study of compactified extra dimensions with appropriate
boundary conditions has revealed alternative symmetry breaking
mechanisms~\cite{Scherk:1978ta,Scherk:1979zr,Hosotani:1983xw,Candelas:1985en,Witten:1985xc,Dixon:1985jw}.
Recent years have seen the development of new ideas in symmetry
reduction which utilise the discrete symmetry transformations of
fields which propagate in compact extra
dimensions~\cite{Kawamura:1999nj}. These methods have been applied to
$SU(5)$ and $SO(10)$ Grand Unified
Theories~\cite{Kawamura:2000ev,Kawamura:2000ir,Altarelli:2001qj,Kobakhidze:2001yk,Hall:2001pg,Hebecker:2001wq,Hebecker:2001jb,Chang:2003gv,Csaki:2001qm,Maru:2001ch,Hall:2001xr,Hall:2002ci,Dermisek:2001hp,Watari:2001pj,Babu:2002ti,PaccettiCorreia:2002xs,PaccettiCorreia:2002eq,Haba:2002ve,Barr:2002fb,Dorsner:2003yg},
left-right symmetric
models~\cite{Mimura:2002te,Mohapatra:2002rn,Perez:2002wb}, the 3-3-1
model~\cite{Gogoladze:2002nq}, unified $SU_W(3)$ models
\cite{Hall:2002rk,Li:2002pb,Dimopoulos:2002jf,Chang:2003sx,Chang:2003rp}
and trinification models~\cite{Carone:2004rp,Carone:2005rk}.

In the present work we study a five dimensional QL symmetric
model. The QL symmetric gauge group is broken by a
combination of orbifold compactification, with the extra dimension
forming an $S^1/Z_2\times Z_2'$ orbifold, and the usual Higgs
mechanism. We assume that fermions are confined to a brane at an
orbifold fixed point, whilst scalar and gauge fields propagate in the
bulk. The physical spectrum of the resulting model is compared with
that of the usual four dimensional models.

We find that the massive
gauge bosons associated with the breaking of leptonic colour
$SU_l(3)\rightarrow SU_l(2)$ appear at two scales. The charge $\pm1/2$
exotic $Y$ bosons do not possess zero Kaluza-Klein modes and thus appear
only at the inverse compactification scale. Direct bounds on
processes mediated by these bosons give
$1/R \ge 5$~TeV. However the ultra-violet (UV) cut-off of the model is
required to be larger than $10^{11}$~GeV to prevent rapid proton
decay. If we assume that the more fundamental theory becomes important
just beyond
the inverse compactification scale we obtain $1/R\gtrsim
10^{9}$~GeV and the $Y$ bosons disappear from the low energy
spectrum. An additional neutral gauge boson, $Z'$, and
exotic charge $1/2$ coloured leptons appear
at a symmetry breaking scale between the inverse compactification
scale and the electroweak scale. This intermediate scale may be as low
as a TeV, rendering these states observable at future colliders. This
differs from four dimensional models where the $Y$ and $Z'$ mass
scales are coupled.

The layout of this paper is as follows. In
Section~\ref{sec:orb_ql_intro_review} we review the QL
symmetric theory. We develop the symmetry breaking used in the five
dimensional theory in Section~\ref{sec:orb_ql_intro_orb_breaking}, and
study the spectrum of exotic gauge bosons and leptons in
Section~\ref{orb_ql_intro_gauge_bosons}. In
Section~\ref{sec:orb_ql_intro_neutrino_mass_p_dcay} we discuss
neutrino mass and proton decay within the model and we conclude in
Section~\ref{orb_ql_intro_conclusion}.
\section{Review Of The Minimal Quark-Lepton Symmetric Model\label{sec:orb_ql_intro_review}}
QL symmetry is implemented by extending the SM gauge group to $\mathcal{G}_{ql}= SU_l(3)\otimes SU_q(3)\otimes SU_L(2)\otimes U_X(1)$, where $SU_q(3)$ is the normal colour group and $X\ne Y$, where $Y$ is the usual SM hypercharge. The matter fields are assigned to the following representations of $\mathcal{G}_{ql}$:
\begin{eqnarray}
& &Q_L\sim (1,3,2,1/3),\nonumber\\
& &u_R\sim (1,3,1,4/3),\mkern20mu d_R\sim(1,3,1,-2/3),\nonumber\\
& &F_L\sim (3,1,2,-1/3),\label{orb_ql_intro_ql_fermi_content}\\
& &E_R\sim (3,1,1,-4/3),\mkern20mu N_R\sim(3,1,1,2/3),\nonumber
\end{eqnarray}
where $F_L$, $N_R$ and $E_R$ contain the usual left-chiral lepton $SU_L(2)$ doublet, right-chiral neutrino and right-chiral charged lepton respectively. The simplest Higgs sector is given by~\cite{Levin:1993sq}
\begin{eqnarray}
& &\chi\sim(3,1,1,2/3),\mkern20mu \chi'\sim(1,3,1,-2/3),\nonumber\\
& &\phi_1\sim(1,1,2,1),\mkern20mu \phi_2\sim(1,1,2,-1).
\end{eqnarray}
The discrete QL symmetry acts on the matter and scalar fields as follows:
\begin{eqnarray}
& & Q_L\leftrightarrow F_L,\mkern20mu u_R\leftrightarrow E_R,\mkern20mu d_R\leftrightarrow N_R,\nonumber\\
& &\chi\leftrightarrow \chi',\mkern20mu \phi_1\leftrightarrow\phi_2,\mkern20mu \phi_1^c\leftrightarrow\phi_2^c,
\end{eqnarray}
where $\phi_{1,2}^c\equiv i\tau_2\phi^*_{1,2}$. The gauge fields transform as:
\begin{eqnarray}
G_q^\mu\leftrightarrow G_l^\mu,\mkern20mu W^\mu\leftrightarrow W^\mu,\mkern20mu C^\mu\leftrightarrow -C^\mu,
\end{eqnarray}
where $G_{q,l}^\mu$ are the $SU_{q,l}(3)$ gauge bosons, $W^\mu$ are the weak bosons and $C^\mu$ is the $U_X(1)$ gauge boson. The Yukawa Lagrangian may be separated into an electroweak portion,
\begin{widetext}
\begin{eqnarray}
\mathcal{L}_{ew}&=&\lambda_1(\overline{Q}_Ld_R\phi_1 +\overline{F}_LN_R\phi_2)+\lambda_1'(\overline{Q}_Ld_R\phi_2^c +\overline{F}_LN_R\phi_1^c)
+\lambda_2(\overline{Q}_Lu_R\phi_1^c+\overline{F}_LE_R\phi_2^c)\nonumber\\
& &+\lambda_2'(\overline{Q}_Lu_R\phi_2 +\overline{F}_LE_R\phi_1)+\mathrm{H.c.},\label{orb_ql_intro_ew_yukawa_Lagrangian}
\end{eqnarray} 
\end{widetext}
and a non-electroweak portion,

\begin{eqnarray}
\mathcal{L}_{non-ew}&=&h_1[\overline{(F_L)^c}F_L\chi +\overline{(Q_L)^c}Q_L\chi']+\nonumber\\
& &h_2[\overline{(E_R)^c}N_R\chi+\overline{(u_R)^c}d_R\chi']+\mathrm{H.c.},
\end{eqnarray}
where $\lambda_{1,2}$, $\lambda'_{1,2}$ and $h_{1,2}$ are Yukawa coupling constants. The scalar potential admits a minimum corresponding to~\cite{Foot:1990dw}
\begin{eqnarray}
& &\langle\phi_1\rangle=(0,u_1)^T,\mkern20mu \langle\phi_2\rangle=(u_2,0)^T,\nonumber\\
& & \langle\chi\rangle=(w,0,0)^T,\mkern20mu \langle\chi'\rangle=0,
\end{eqnarray}
where an $SU_{l}(3)$ rotation has been performed to obtain the most general vacuum expectation value (VEV) for $\chi$. The vanishing VEV for $\chi'$ ensures that $SU_q(3)$ remains unbroken whilst the non-zero VEV for $\chi$ breaks the gauge group as
\begin{eqnarray}
\mathcal{G}_{ql}\rightarrow SU_l(2)\otimes SU_q(3)\otimes SU_L(2)\otimes U_Y(1).
\end{eqnarray}
Here the SM hypercharge generator is given by:
\begin{eqnarray}
Y=X+\frac{1}{\sqrt{3}}T_8,
\end{eqnarray}
where $T_8=(1/\sqrt{3})\times\mathrm{diag}(-2,1,1)$ is a diagonal generator of $SU_l(3)$. At the next stage of symmetry breaking the SM gauge group is broken by the non-zero VEV's for $\phi_{1,2}$.
\begin{eqnarray}
SU_l(2)\otimes\mathcal{G}_{SM}\rightarrow SU_l(2)\otimes SU_q(3)\otimes U_Q(1),
\end{eqnarray}
where $\mathcal{G}_{SM}$ is the SM gauge group and $U_Q(1)$ is the electromagnetic gauge group. Note that an $SU_l(2)$ subgroup of $SU_l(3)$ remains unbroken.

At the first stage of symmetry breaking the non-zero VEV for $\chi$ breaks five generators of $\mathcal{G}_{ql}$ and thus five gauge bosons develop masses of order $w$. This amounts to four gauge bosons with charge 1/2 and one neutral gauge boson (which mixes of course with the other neutral gauge bosons). The additional lepton degrees of freedom form vectorial representations of the unbroken subgroup and also acquire masses of order $w$. All SM fermions remain massless at this stage, forming chiral representations of the remaining symmetry. At the second stage of symmetry breaking the SM gauge group is broken by the non-zero VEV's $u_{1,2}$ and the SM fermions acquire Dirac mass terms. The electroweak Yukawa Lagrangian (\ref{orb_ql_intro_ew_yukawa_Lagrangian}) ensures that all SM fermions may have unique masses with no relations of the type $m_e=m_u$ or $m_e=m_d$ arising (the field $\phi_2$ is included for this very purpose). The photon remains massless whilst the other SM gauge bosons develop order $u_{1,2}$ masses.

Thus, to surmise, the model predicts an additional massive neutral gauge boson $Z'$ and four charge 1/2 massive gauge bosons, which we generically label as $Y$,  all with order $w$ masses. It also predicts additional leptons, forming non-trivial representations of the unbroken $SU_l(2)$, with order $w$ masses. Phenomenology of these additional states can be found in~\cite{Foot:1991fk}. 
\section{Orbifold Reduction Of Quark-Lepton Symmetry\label{sec:orb_ql_intro_orb_breaking}}
Having reviewed the standard four dimensional QL symmetric model we now develop a five dimensional version of the model. The additional spatial dimension is taken as the orbifold $S^1/Z_2\times Z_2'$, whose coordinate is labelled as $y$. The construction of the orbifold proceeds via the identification $y\rightarrow -y$ under the $Z_2$ symmetry and $y'\rightarrow-y'$ under the $Z_2'$ symmetry, where $y'=y+\pi R/2$. The physical region in $y$ is given by the interval $[0,\pi R/2]$. We take the same particle content as the previous section. All gauge bosons and scalar fields are assumed to propagate in the bulk, whilst the matter fields are confined to a four dimensional wall at the orbifold fixed point $y=0$. The bulk gauge group $\mathcal{G}_{ql}$ is reduced by the orbifold compactification. The five dimensional Lagrangian is invariant under the discrete $Z_2\times Z_2'$ symmetry, which acts on the gauge bosons as follows:
\begin{eqnarray}
W_\mu(x^\mu,y)&\rightarrow& W_\mu(x^\mu,-y)=PW_\mu(x^\mu,y)P^{-1},\nonumber\\
W_5(x^\mu,y)&\rightarrow& W_5(x^\mu,-y)=-PW_5(x^\mu,y)P^{-1},\nonumber\\
W_\mu(x^\mu,y')&\rightarrow& W_\mu(x^\mu,-y')=P'W_\mu(x^\mu,y')P'^{-1},\nonumber\\
W_5(x^\mu,y')&\rightarrow& W_5(x^\mu,-y')=-P'W_5(x^\mu,y')P'^{-1}.\nonumber
\end{eqnarray}We take $P$ and $P'$ to be trivial for the $SU_q(3)$, $SU_L(2)$ and $U_X(1)$ gauge bosons. For the $SU_l(3)$ gauge bosons we choose $P=\mathrm{diag}(1,1,1)$ and $P'=\mathrm{diag}(-1,1,1)$. We write the five dimensional $SU_l(3)$ gauge bosons as
\begin{eqnarray}
G_l&=&T_a G_l^a \nonumber\\
&=&\left( \begin{array}{ccc}
           -\frac{2}{\sqrt{3}}G^8&\sqrt{2}Y^1 &\sqrt{2}Y^2    \\
           \sqrt{2}Y^{1\dagger}&G^3+\frac{1}{\sqrt{3}}G^8& \sqrt{2}\tilde{G}\\
           \sqrt{2}Y^{2\dagger}&\sqrt{2}\tilde{G}^\dagger&-G^3+\frac{1}{\sqrt{3}}G^8  
\end{array} \right),
\end{eqnarray}
and find their $Z_2\times Z_2'$ parities to be
\begin{eqnarray}
Y_{\mu}^1,Y_{\mu}^2,Y_{\mu}^{1\dagger},Y_{\mu}^{2\dagger}&\rightarrow& (+,-),\nonumber\\
Y_{5}^1,Y_{5}^2,Y_{5}^{1\dagger},Y_{5}^{2\dagger}&\rightarrow& (-,+),\nonumber\\
G^8_\mu,G^3_\mu,\tilde{G}_\mu,\tilde{G}_\mu^\dagger&\rightarrow& (+,+),\nonumber\\
G^8_5,G^3_5,\tilde{G}_5,\tilde{G}_5^\dagger&\rightarrow& (-,-).
\end{eqnarray}
The compact fifth dimension allows one to expand the gauge bosons as a Fourier series, with the $Z_2\times Z_2'$ parities constraining the series as usual.
\begin{widetext}
\begin{eqnarray}
\psi_{(+,+)}(x^\mu,y)&=&\frac{2}{\sqrt{\pi R}}\left(\frac{1}{\sqrt{2}}\psi^{(0)}_{(+,+)}(x^\mu)+\sum_{n=1}^\infty\psi^{(n)}_{(+,+)}(x^\mu)\cos\frac{2ny}{R}\right),\nonumber\\
\psi_{(+,-)}(x^\mu,y)&=&\frac{2}{\sqrt{\pi R}}\sum_{n=1}^\infty\psi^{(n)}_{(+,-)}(x^\mu)\cos\frac{(2n-1)y}{R},\nonumber\\
\psi_{(-,+)}(x^\mu,y)&=&\frac{2}{\sqrt{\pi R}}\sum_{n=1}^\infty\psi^{(n)}_{(-,+)}(x^\mu)\sin\frac{(2n-1)y}{R},\nonumber\\
\psi_{(-,-)}(x^\mu,y)&=&\frac{2}{\sqrt{\pi R}}\sum_{n=1}^\infty\psi^{(n)}_{(-,-)}(x^\mu)\sin\frac{2ny}{R},
\end{eqnarray}
\end{widetext}
where $\psi$ represents a generic field. Thus the four dimensional charge 1/2 bosons $Y_{\mu}^1$, $Y_{\mu}^2$, $Y_{\mu}^{1\dagger}$ and $Y_{\mu}^{2\dagger}$ do not possess zero modes, with the $n$th mode possessing a mass of $(2n-1)/R$. The fields $G^8_\mu$, $G^3_\mu$, $\tilde{G}_\mu$ and $\tilde{G}_\mu^\dagger$ all have zero modes, with the higher modes possessing a mass of $2n/R$. We see that the bulk symmetry $SU_l(3)$ has been reduced to $SU_l(2)\otimes U_{X'}(1)$ at the zero mode level. This is analogous to the $SU(3)$ symmetry reduction employed in~\cite{Gogoladze:2002nq,Carone:2005rk}. After compactification the zero mode gauge group is
\begin{eqnarray}
& &SU_l(2)\otimes SU_q(3)\otimes SU_L(2)\otimes U_{X'}(1)\otimes U_X(1)\nonumber
\end{eqnarray}
and the next stage of symmetry breaking requires
\begin{eqnarray}
U_{X'}(1)\otimes U_X(1)\rightarrow U_Y(1),
\end{eqnarray}
which shall be achieved by the usual Higgs mechanism. The $Z_2\times Z_2'$ parities of the $\phi_{1,2}$ are
\begin{eqnarray}
\phi_{1,2}(x^\mu,y)&\rightarrow& \phi_{1,2}(x^\mu,-y)=P\phi_{1,2}(x^\mu,y),\nonumber\\
\phi_{1,2}(x^\mu,y')&\rightarrow& \phi_{1,2}(x^\mu,-y')=P'\phi_{1,2}(x^\mu,y'),
\end{eqnarray}
where $P=P'=\mathrm{diag}(1,1)$. For $\chi$ we have
\begin{eqnarray}
\chi(x^\mu,y)&\rightarrow& \chi(x^\mu,-y)=P\chi(x^\mu,y),\nonumber\\
\chi(x^\mu,y')&\rightarrow& \chi(x^\mu,-y')=-P'\chi(x^\mu,y'),\label{orb_ql_intro_chi_parities}
\end{eqnarray}
with $P=\mathrm{diag}(1,1,1)$ and $P'=\mathrm{diag}(-1,1,1)$, whilst for $\chi'$ we take
\begin{eqnarray}
\chi'(x^\mu,y)&\rightarrow& \chi'(x^\mu,-y)=-P\chi'(x^\mu,y),\nonumber\\
\chi'(x^\mu,y')&\rightarrow& \chi'(x^\mu,-y')=P'\chi'(x^\mu,y'),\label{orb_ql_intro_chi'_parities}
\end{eqnarray}
with $P$ and $P'$ trivial. Thus $\chi'\rightarrow(-,+)$ and vanishes
at the $y=0$ boundary of the extra dimension. Under the symmetry reduction
\begin{eqnarray}
SU_l(3)\rightarrow SU_l(2)\otimes U_{X'}(1),
\end{eqnarray} 
one has
\begin{eqnarray}
\chi\rightarrow \chi_2 \oplus \chi_1,
\end{eqnarray}
where $\chi_2\sim(2,1)$ and $\chi_1\sim(1,-2)$ have the $Z_2\times Z_2'$ parities
\begin{eqnarray}
\chi_1\rightarrow (+,+)\mkern15mu,\mkern15mu\chi_2\rightarrow(+,-).
\end{eqnarray}
The zero mode for $\chi_1$ may develop a VEV and break the gauge symmetry as follows:
\begin{widetext}
\begin{eqnarray}
SU_l(2)\otimes SU_q(3)\otimes SU_L(2)\otimes U_{X'}(1)\otimes U_X(1)\rightarrow SU_l(2)\otimes SU_q(3)\otimes SU_L(2)\otimes U_Y(1).
\end{eqnarray}
\end{widetext}
The final stage of symmetry breaking occurs when the neutral components of $\phi_{1,2}$ develop the VEV's $u_{1,2}$ to give
\begin{eqnarray}
& &SU_l(2)\otimes SU_q(3)\otimes SU_L(2)\otimes U_Y(1)\nonumber\\
& &\mkern80mu\downarrow\nonumber\\
& &SU_l(2)\otimes SU_q(3)\otimes U_Q(1).
\end{eqnarray}
\section{Gauge Bosons And Exotic Leptons\label{orb_ql_intro_gauge_bosons}}
In the previous section we have shown how a combination of orbifold
compactification and the Higgs mechanism may be used to break a five
dimensional QL symmetric model down to something resembling the SM. In
this section we shall discuss the mass scale of the gauge bosons in
the model outlined. We denote the VEV's of the scalars as
\begin{eqnarray}
\langle\chi_1\rangle &=&w\sqrt{2/\pi R},\nonumber\\
\langle\phi^0_{1,2}\rangle&=&u_{1,2}\sqrt{2/\pi R},\nonumber
\end{eqnarray}
and we also define $u^2=u_1^2+u^2_2$. In the basis
$(W^{0(n)},C^{(n)},G^{8(n)})$ the mass squared matrix for the neutral
gauge bosons is
\begin{widetext}
\begin{eqnarray}
\mathcal{M}^{2(n)}_{neutral}=\left( \begin{array}{ccc}
           g_L^2\frac{u^2}{2}+\left(\frac{2n}{R}\right)^2& -g_Xg_L\frac{u^2}{2}&0  \\ -g_Xg_L\frac{u^2}{2}& g_X^2\frac{u^2}{2}+\frac{2}{9}g_X^2w^2+\left(\frac{2n}{R}\right)^2&\frac{2\sqrt{3}}{9}g_Xg_Sw^2\\0&\frac{2\sqrt{3}}{9}g_Xg_Sw^2&\frac{2}{3}g_S^2w^2+\left(\frac{2n}{R}\right)^2
\end{array} \right),
\end{eqnarray}
\end{widetext}
where $g_L$ [$g_X$] is the $SU_L(2)$ [$U_X(1)$] coupling constant and
$g_S$ denotes the common $SU_l(3)$ and $SU_q(3)$ coupling
constant. Note that these are dimensionless constants, related to the
dimensionfull five dimensional Lagrangian constants via
$g_S=g_S^5\sqrt{2/\pi R}$ etc. Assuming $g_S^2\gg g_X^2,g_L^2$ one may
write the eigenvalues as
\begin{eqnarray}
M^{2(n)}_{\gamma}&=&\left(\frac{2n}{R}\right)^2,\nonumber\\
M_{Z}^{2(n)}&\simeq&\frac{1}{2}(g_X^2+g_L^2)u^2-\frac{g_X^2}{6g_S^2}u^2+\left(\frac{2n}{R}\right)^2,\nonumber
\end{eqnarray}
\begin{eqnarray}
M_{Z'}^{2(n)}&\simeq&\frac{2}{3}g_S^2w^2\left\{1+\frac{g_X^2}{3g_S^2}\right\}+\left(\frac{2n}{R}\right)^2.
\end{eqnarray}
Note that the zero modes possess the same eigenvalues as the neutral gauge bosons in the minimal four dimensional QL symmetric model~\cite{Foot:1990dw,Foot:1991fk}. In fact these zero modes couple to fermions in exactly the same way as the neutral gauge bosons in the minimal QL symmetric model, making the phenomenology of these sates identical to that of the neutral gauge bosons studied in~\cite{Foot:1991fk}.

The zero modes consist of the massless photon, the $Z$ boson with mass of order $u$, the electroweak scale, and an additional neutral boson $Z'$ with mass of order $w$, the $U_{X'}(1)\otimes U_X(1)$ symmetry breaking scale. The phenomenological bound of $M_{Z'}>720$~GeV obtained in~\cite{Foot:1991fk} also applies to the zero mode $Z'$ boson in the present model. Thus we obtain a lower bound on the $U_{X'}(1)\otimes U_X(1)$ symmetry breaking scale of $w\gtrsim1$~TeV, which is low enough to permit observation of the $Z'$ boson and the exotic leptons at the LHC.

The mass of the $W$ bosons in the present model is
\begin{eqnarray}
M^{2(n)}_W=\frac{1}{2}g_L^2u^2 +\left(\frac{2n}{R}\right)^2,
\end{eqnarray}
with the zero mode corresponding to the usual $W$ bosons. The mass of the charged 1/2 bosons is
\begin{eqnarray}
M_{Y^1}^{2(n)}=M_{Y^2}^{2(n)}=\frac{1}{2}g_S^2w^2+\left(\frac{2n-1}{R}\right)^2,\end{eqnarray}
with the zero mode absent. Thus we see that for $1/R>w$ the mass of
the $Y$ bosons is set by the size of the extra
dimensions. In~\cite{Foot:1991fk} a rough lower bound on the mass of
the $Y$ bosons was obtained by considering the rare decay
$\mu\rightarrow 3e$, which is radiatively induced in QL symmetric
models due to the presence of the exotic leptons. The result is
\begin{eqnarray}
M_{Y}\ge (5\ \mathrm{TeV})\times c,
\end{eqnarray}
where $c$ is an $O(1)$ number which depends on mixing angles and the
mass of the exotic leptons. In the present model this translates into
the rough bound
\begin{eqnarray}
\frac{1}{R}\ge 5\ \mathrm{TeV}\label{orb_ql_intro_bound_r}.
\end{eqnarray}
Note that the present model provides a rationale for the hierarchy $M_{Z'}<M_{Y}$ with the mass of these bosons originating from different symmetry breaking mechanisms. The $Y$ bosons, possessing no zero mode, acquire mass at the inverse compactification scale $1/R$, whilst the lightest $Z'$ obtains mass at the $U_{X'}(1)\otimes U_X(1)$ symmetry breaking scale via the Higgs mechanism.
 
The exotic charge 1/2 leptons, known as liptons in the literature~\cite{Foot:1991fk},
also develop mass at the
$U_{X'}(1)\otimes U_X(1)$ symmetry breaking scale $w$. These states
are confined by the unbroken $SU_l(2)$ symmetry and form exotic hadrons,
an analysis of which may be found in~\cite{Foot:1991fk} and more
recently in~\cite{Foot:2006ie} (see also~\cite{Babu:2003nw}). Here we briefly
 surmise some of the interesting phenomenology associated with these
 states.

The liptons are confined into two particle bound states by the
unbroken $SU_l(2)$ symmetry. The bound states formed by the lightest
lipton (which we call $L_1$) can be produced by virtual $W$, $Z$ and
$\gamma$ decays in future colliders such as the LHC. The $SU_l(2)$ gauge
interactions preserve a global flavour symmetry $SU_F(2)$ with $L_1$
and its antiparticle $L_1^c$ forming a doublet representation of $SU_F(2)$. The exotic hadron
flavour structure can be determined by the group product
\begin{eqnarray}
2\otimes 2=1_A\oplus 3_S,
\end{eqnarray}
where the subscripts indicate the symmetry properties of the state
under interchange of the liptons. The ground states of the exotic
hadrons have zero angular momentum and thus the Pauli principle
dictates that the flavour triplet will have spin-1 and the flavour
singlet will have spin-0. These states may be denoted as
\begin{eqnarray}
\rho^+&=& L_1L_1,\mkern15mu \rho^-=L_1^cL_1^c,\nonumber\\
\rho^0&=&\frac{1}{\sqrt{2}}
(L_1L_1^c+L_1^cL_1),\nonumber\\
\xi^0&=&\frac{1}{\sqrt{2}}
(L_1L_1^c-L_1^cL_1),
\end{eqnarray}
and all decay into SM particles, with decay modes such as
\begin{eqnarray}
\rho^+&\rightarrow& e^+\nu,\nonumber\\
\rho^0&\rightarrow& e^+e^-,\nonumber\\
\xi^0&\rightarrow&2\gamma.
\end{eqnarray}

Let us now stop to contrast the exotic particle spectrum in our model with
that of the minimal QL symmetric model. In the minimal QL symmetric
model the exotic bosons and fermions all develop mass at the scale
$w$. The slightly more stringent bound on the $Y$ bosons requires this
scale to be larger than a TeV or so. Our five dimensional model
decouples the scale of some of these exotic particles. The $Y$ bosons
develop mass at the inverse compactification scale due to the absence
of zero modes. The additional neutral boson $Z'$ and the
liptons develop mass at the $U_{X'}(1)\otimes U_X(1)$ symmetry
breaking scale, which is less than the inverse compactification
scale. At the scale $M_Y$ the first Kaluza-Klein modes of the photon,
the $Z$ and the $Z'$ also appear, providing a clear distinction
between the higher dimensional model and the minimal four dimensional
model.

We note that we have considered the minimal Higgs sector to date. Some studies of QL symmetric models have included the additional scalars~\cite{Foot:1990dw,Foot:1991fk}
\begin{eqnarray}
\Delta_l\sim(\bar{6},1,1,-4/3),\mkern20mu \Delta_q\sim(1,\bar{6},1,4/3),
\end{eqnarray}
which form a pair under the discrete QL symmetry. If
$\langle\Delta_q\rangle=0$ and the component of $\Delta_l$ which forms
a $(1,4)$ representation of $SU_l(2)\otimes U_{X'}(1)\subset SU_l(3)$
develops a non-zero VEV $\langle\Delta_l\rangle$, the QL
symmetric gauge group is broken down to $SU_l(2)\otimes SU_q(3)\otimes
SU_L(2)\otimes U_Y(1)$ at the scale $\langle\Delta_l\rangle$. This
extension has the advantage of giving a Majorana mass to the
right-chiral neutrinos at the scale $\langle\Delta_l\rangle$, allowing
one to employ the see-saw mechanism to explain the relative lightness
of the neutrinos.

In the minimal QL symmetric model one requires highly tuned
Dirac Yukawa couplings to produce very light neutrinos, making the
addition of the states $\Delta$ an attractive extension. However this
modification couples the mass of the exotic gauge bosons $Y$ and $Z'$
to the right-chiral Majorana mass scale, which is typically required
to be larger than $10^{11}$~GeV or so, depending on how small one
tolerates the neutrino Dirac mass matrix Yukawa couplings. Thus the
exotic bosons become unobservably heavy. The liptons do not
develop mass at the scale $\langle\Delta_l\rangle$ and one still
requires the field $\chi$ to develop a VEV (though it is not breaking
any symmetry) to give mass to these fermions.

Thus in the extended four dimensional QL symmetric model the
liptons have masses of order $w$ and the exotic gauge bosons
have mass at the scale $\langle\Delta_l\rangle$. Note that the
hierarchy $\langle\Delta_l\rangle \gg w$ is still allowed in these
models so that the exotic fermions may be observed at TeV
energies. Again this spectrum contrasts vividly with that obtained in
the five dimensional model. In the four dimensional model only the
liptons become accessible at the scale $w$, whilst the five
dimensional model also requires the $Z'$ boson to appear at this
scale.
\section{Fermions, Proton
 Decay and Neutrino Mass\label{sec:orb_ql_intro_neutrino_mass_p_dcay}}
In the present work we shall assume that the SM fermions are confined
to the brane at $y=0$. We have seen in the previous sections that the
mass scale of the $Y$ bosons is set by the inverse compactification
scale in our framework, which is bound to be greater than 5~TeV. Given
that all fermions are
assumed to be stuck
at the $y=0$ wall, non-renormalizable proton decay inducing operators
will arise and we must ensure that these are adequately suppressed. 

In the SM proton decay occurs via the dimension six operator
\begin{eqnarray}
\frac{h}{\Lambda_{SM}^2}\epsilon_{\alpha\beta\gamma}Q^\alpha Q^\beta Q^\gamma L,\label{orb_ql_intro_p_decay_operator}
\end{eqnarray}
where \mbox{$\alpha,\beta,\gamma=1,2,3$} are colour indices, $\Lambda_{SM}$ is the SM UV cutoff, $Q$ ($L$) denotes quark (lepton) operators and $h$ is a dimensionless coupling. This operator leads to the decay $p\rightarrow e^+\pi^0$. Given that experimental bounds require the lifetime of the proton to be in excess of $1.6\times10^{33}$~years, one requires $\Lambda_{SM}\sim 10^{16}$~GeV~\cite{Nath:2006ut}. In a four dimensional QL symmetric framework the lowest dimension non-renormalizable operator that induces proton decay has dimension seven,
\begin{eqnarray}
\frac{h}{\Lambda_{ql}^3}\epsilon_{\alpha\beta\gamma}Q^\alpha Q^\beta Q^\gamma \chi^{\dagger}_{\bar{\alpha}}L^{\bar{\alpha}},\label{orb_ql_intro_4d_ql_pdecay}
\end{eqnarray}
where $\bar{\alpha}=1,2,3$ is the leptonic colour index and $\Lambda_{ql}$ is the UV cutoff. When $\chi$ develops a VEV this leads to an effective operator of the form (\ref{orb_ql_intro_p_decay_operator}). Relating the two cutoff's gives
\begin{eqnarray}
\Lambda_{ql}=\Lambda_{SM}^{2/3}\langle\chi\rangle^{1/3}.
\end{eqnarray}
and using the order TeV lower bound on $\langle\chi\rangle$ leads to
\begin{eqnarray}
\Lambda_{ql}\ge 5\times 10^{11}~\mathrm{GeV}.
\end{eqnarray} 

If quarks and leptons are localized at $y=0$ in our five
dimensional QL model the lowest dimension non-renormalizable operator
which leads to proton decay is the equivalent of
(\ref{orb_ql_intro_4d_ql_pdecay}):
\begin{eqnarray}
\frac{h}{\Lambda_{ql}^{7/2}}\epsilon_{\alpha\beta\gamma}Q^\alpha Q^\beta Q^\gamma \chi^{\dagger}_{\bar{\alpha}}L^{\bar{\alpha}},\label{orb_ql_intro_p_decay_operator_5d}
\end{eqnarray}
where $\chi$ is now a five dimensional field.
If we assume that the cut-off is only a couple of orders of magnitude
larger than the inverse compactification scale, say $\Lambda_{ql}\sim
100/R$, we arrive at the bound
\begin{eqnarray}
\Lambda_{ql}\ge 2\times 10^{11}~\mathrm{GeV},
\end{eqnarray}
which is incompatible with a low fundamental scale.
The bound on $\Lambda_{ql}$ corresponds to an inverse
compactification scale of order $10^{9}$~GeV and consequently the phenomenology associated with the
$Y$ bosons will not appear at low energies within in our
framework. The bounds from the proton decay rate do not disturb the
bounds on the $Z'$ boson mass obtained earlier and this boson can, in
principle, still appear at order TeV energies.

As mentioned in the previous section, it is known that in a four
dimensional QL symmetric model one may suppress the known neutrino
masses relative to the electroweak scale by introducing scalars
forming a six dimensional representation of $SU_l(3)$ (and their QL
symmetry partners). However this
forces the $Z'$ mass up to the right-chiral neutrino Majorana mass
scale. In this section we
shall employ a mechanism previously employed in QL
symmetric models~\cite{Foot:1995xx} to allow the neutrinos to acquire
masses suppressed relative to the electroweak scale
(see~\cite{Wyler:1982dd,Mohapatra:1986ks} for earlier implementations
of this mechanism). The mechanism
does not require the introduction
of any additional high energy scales and allows the zero mode $Z'$ boson to
retain an order TeV mass.

We add three fermions $S_L\sim(1,1,1,0)$ to the particle spectrum,
which transform trivially under the discrete QL symmetry. These
fermions are also assumed to be stuck to the wall at $y=0$. The gauge
symmetries allow the additional Lagrangian terms
\begin{eqnarray}
\mathcal{L}_{S_L}=h_3^5[\overline{S}_L\chi^\dagger N_R+\overline{S}_L\chi'^{\dagger}d_R] + M_S\overline{S_L^c}S_L+\mathrm{H.c.},\label{intro_orb_ql_s_terms}
\end{eqnarray}
where we omit the delta function indicating that the fermions are
localized at the $y=0$ brane.

After symmetry breaking is complete the non zero VEV's for $\phi_{1,2}$ and $\chi$ result in the neutrino mass matrix
\begin{eqnarray}
\mathcal{M}_\nu=
\left(\begin{array}{ccc} 0 & m & 0 \\m &
0 & M\\0 & M & M_{S}
\end{array}\right),\label{orb_intro_double_see_saw_mass_matrix}
\end{eqnarray}
in the Majorana basis $(\nu_L,(\nu_R)^c,S_L)$, where for simplicity we show only one generation. Here $\nu_L$ ($\nu_R$) is a normal electroweak doublet (singlet) neutrino whilst
\begin{eqnarray}
m=\lambda_1u_2+\lambda_1'u_1^*\sim u_{1,2}\mkern10mu \mathrm{and}\mkern10mu M=h_3w,
\end{eqnarray}
where $h_3=h_3^5\sqrt{2/\pi R}$. In the limit $M_S\rightarrow 0$ the mass eigenstates consist of one massless Weyl neutrino
\begin{eqnarray}
\nu_{WL}=\cos\theta\nu_L-\sin\theta S_L,
\end{eqnarray}
and one Dirac neutrino with mass $\sqrt{m^2+M^2}$,
\begin{eqnarray}
\nu_{DL}&=&\sin\theta\nu_L+\cos\theta S_L,\nonumber\\
\nu_{DR}&=&\nu_R,
\end{eqnarray}
where $\tan\theta=m/M$. When $M_S$ is turned on, and the hierarchy
$M_S\ll m\ll M$ exists, the massless state develops a mass
$M_Sm^2/M^2$ and the Dirac neutrino becomes two Majorana neutrinos
with mass splitting of order $M_S$. Note that the light state consists
predominantly of the normal electroweak neutrino $\nu_L$ under the
hierarchy $m\ll M$. If one takes $M\sim1$~TeV, in line with the
lower bound on the $Z'$ boson masses, and $m\leq 10$~GeV, then
$\theta \leq 0.6^\circ$. Requiring $M_Sm^2/M^2\sim0.1$~eV to accommodate
solar and atmospheric neutrino oscillation data then gives
$M_S\sim1$~keV. As noted in~\cite{Foot:1995xx}, the hierarchy between
the scale $M_S$ and $u_{1,2}$, $w$ is perturbatively stable due to the
enhanced symmetry of the theory in the limit $M_S\rightarrow 0$.

The three generation case is readily obtained by converting $m$, $M$
and $M_S$ to $3\times3$ matrices in flavour space. The result been
three mixed, light neutrinos, with mass of order $M_sm^2/M^2$, which
predominantly contain the usual electroweak neutrinos $\nu_{eL}$,
$\nu_{\mu L}$ and $\nu_{\tau L}$.

Note that in a four dimensional QL model the Lagrangian term $S_L\chi'd$ in
eq. (\ref{intro_orb_ql_s_terms}) breaks baryon number and allows the proton
to decay. However in our five dimensional model the $Z_2\times
Z_2'$ parities for $\chi'$ ensure that it vanishes at the $y=0$
brane. Thus after integrating over the extra dimension this term
disappears from the four dimensional theory and baryon number is
conserved. Consequently proton decay may only be mediated by heavy
fields in the more fundamental theory.

This mechanism of obtaining light neutrino masses is particularly well
suited to higher dimensional theories when the right-chiral neutrinos
form non-trivial representations of some additional gauge
symmetry. Provided the additional symmetry is broken above the
electroweak scale one may obtain light neutrinos with only a mild
hierarchy between the electroweak scale and the additional symmetry
breaking scale. Whilst we have used this mechanism in a QL symmetric
model it could also be applied in, $eg$ a higher dimensional
left-right symmetric framework (examples of which
include~\cite{Mimura:2002te,Mohapatra:2002rn,Perez:2002wb}).

We note that it may be possible to construct a five dimensional
QL symmetric model compatible with a low fundamental scale
by allowing the fermions to propagate in the bulk. One could then
localize quarks and leptons at different points in the extra
dimension~\cite{Arkani-Hamed:1999dc}, thereby reducing their fifth
dimensional wavefunction overlaps and suppressing the proton decay rate. This idea is currently under investigation~\cite{intro_orb_ql_in_prep}
\section{Conclusion\label{orb_ql_intro_conclusion}}
We have studied the five dimensional QL symmetric gauge group. The $SU_l(3)$
symmetry was reduced to $SU_l(2)\otimes U_{X'}(1)$ as a result of the
compactification of the fifth dimension. The charged 1/2 $Y$ bosons,
corresponding to the broken $SU_l(3)$ generators, form a Kaluza-Klein
tower with no zero mode. Consequently the lightest $Y$ bosons
possesses a mass of order $1/R$, with direct bounds giving $1/R\ge 5$~TeV.
However the prevention of rapid proton decay requires a UV cut-off of
order $10^{11}$~GeV. If one assumes that the more fundamental theory
takes over just beyond the inverse compactification scale one expects $1R\gtrsim
10^{9}$~GeV, and the $Y$ bosons disappear from the low energy
spectrum.

The subsequent breaking of
$U_{X'}(1)\otimes U_X(1)$ by the
Higgs mechanism produces a massive neutral gauge boson at a scale
phenomenologically required to be greater than $1$~TeV. The
liptons also acquire mass at this stage of symmetry breaking. The
photon, $Z$ and $W$ bosons form part of a Kaluza Klein tower, with the
$n=0$ modes corresponding to the usual SM gauge bosons. The bounds on
the exotic
states are low enough for the zero mode of the  additional neutral
gauge boson to appear at an $e^+e^-$ collider operating at TeV
energies and for the exotic leptons to appear at the LHC.

This is to be contrasted with the four dimensional QL symmetric model,
where the massive gauge bosons not contained within the SM acquire
mass at a common symmetry breaking scale. This scale is required to be
of order $10^{11}$~GeV or larger if the see-saw mechanism is employed,
rendering the exotic gauge bosons unobservable at low energies.

We have also shown that one may obtain light neutrinos within the five
dimensional framework without introducing any additional high energy
scales, and although our model is not compatible with a low fundamental
scale this situation may change if the fermions are assumed to propagate
in the bulk~\cite{intro_orb_ql_in_prep}.
\section*{Acknowledgements}
KM thanks T.~Gherghetta, R.~Volkas, A.~Demaria, D.~George,
A.~Coulthurst and R. Foot for
clarifying communications. This
work was supported in part by the Australian Research Council.

\end{document}